\documentclass[preprint,prb,nobibnotes,showkeys,preprintnumbers,amsmath,amssymb]{revtex4}
\usepackage{graphicx}
\usepackage{pst-all}
\usepackage{dcolumn}
\usepackage{bm}
\usepackage[dvips]{epsfig}

\newcommand{\eg}{e.\,g.\ }
\newcommand{\ie}{i.\,e.\ }

\begin{document}
\title{Feynman Path Integral of a charged anisotropic HO\\
in crossed electric and magnetic fields.\\
Alternative calculational methods. 
}
\author{Cyril Belardinelli}
\email[Electronic mail: ]{cyril.belardinelli@hslu.ch}
\affiliation{Lucerne University of Applied Sciences and Arts, Technikumstrasse 21, CH-6048 Horw, Switzerland}
\date{\today}
\begin{abstract}
In the present paper the author evaluates the path integral of a charged anisotropic Harmonic Oscillator (HO) in crossed electric and magnetic fields by two alternative methods. Both methods enable a rather formal calculation and circumvent some mathematical delicate issues such as the occurrence of an infinite Normalization constant and ambiguities with path integral calculations when magnetic fields are present.\\
The \emph{$1^{\text{st}}$} method uses  complex Fourier series and a regularization scheme via the Riemann-$\zeta$-function.
The \emph{$2^{\text{nd}}$} method evaluates the path integral by transforming the Lagrangian to a uniformly rotating system. The latter method uses the fact that the Lorentz- and Coriolis force have the same functional form. 
Both forces cancel each other within the rotating system given that it rotates with Larmor frequency $\omega_{L}$. This fact simplifies considerably the calculation of the path integral.
\end{abstract}


\maketitle
\section{\label{intro}Introduction}
Path integrals are an indispensable computational tool in many areas of theoretical physics. Still today it is unclear how one can make full sense (according to standards of mathematical rigour) of the functional measure 
$\mathcal{D}[\bold{r}(t)]$, see Eq.~(\ref{propagator}). As is well known, a $\sigma$-additive Lebesque-type measure which is translationally and rotationally invariant does not exist in an Hilbert space of infinite dimensions. (See \eg Ref.\cite{mazzucchi:09}). Nevertheless, various attempts have been made at defining it in a more rigorous manner.\cite{mazzucchi:09,kac:51,gelfand:60, montroll:52, cameron:60}Despite the lack of rigour its calculational power is remarkable. The actual situation bears some resemblance to the late $17^{\text{th}}$ century when Leibniz' infinitesimals\cite{leibniz:1684} were applied successfully to various physico-mathematical issues despite their (at the time) rather obscure and mathematically ill-defined status. (See \eg Ref.\cite{kline:1972})\\ 
This paper presents two alternative techniques for evaluating two-dimensional (2D) Gaussian path integrals, \ie path integrals with a quadratic Lagrangian in 2D (to be understood in a general sense, including velocity dependent terms such as magnetic interactions and static electric fields) and aims at symplifying calculations which are rather lengthy when evaluated by standard methods. The main purpose of the paper is to show that the calculation of these path integrals by the presented methods can be done in a rather formal way. 
\\The \emph{$1^{\text{st}}$} method uses \emph{complex} Fourier series and a regularization scheme via the Riemann-$\zeta$-function. The evaluation of path integrals by a \emph{real} Fourier Sine Series is a frequently used method\cite{feynman:2010} where quantum fluctuations around the classical path are expanded in terms of normalised eigenmodes $\phi_{n}(t)$ of the form:
\begin{equation}
\label{real_expansion} 
\begin{split}
\phi_{n}(t):=\sqrt{\frac{2}{T}}\sin{\frac{n\pi}{T}t},  \quad n\in \mathbb{N}
\end{split}
\end{equation}
However, this method has its limitations; it fails when one calculates the path integral for Lagrangians with a velocity dependent potential, such as the magnetic interaction. See App.\eqref{failure_real_fourier}. Indeed, a direct evaluation of such propagators by Fourier expansion is to the author's knowledge absent in the literature. However, in Sec.\eqref{3B} we will see how this important shortcoming can elegantly be resolved if one uses an expansion in terms of complex Fourier series. 
A further advantage of the method is that one can circumvent the need of introducing a (formally divergent) Normalization constant since the applied $\zeta-$regularization leads straightforwardly to the correct propagator.\\The \emph{$2^{\text{nd}}$} method evaluates the path integral by transforming the Lagrangian to a uniformly rotating system. This method uses the fact that the Lorentz force ($\bold{F_{L}}=q\bold{v}\times\bold{B}$) and the Coriolis force ($\bold{F_{c}}=-2m \bold{\omega\times {v}}$) have the same functional form. Both forces cancel each other within the rotating system given that it rotates with Larmor frequency $\omega_{L}=qB/2m$. This method is useful whenever a magnetic field (uniformly or not) is present. Through the rotation the magnetic field is effectively transformed away and one is left with the Lagrangian of an effective harmonic oscillator. The method allows a straightforward calculation of the classical action by deriving it directly from that one of the Harmonic Oscillator. This is a considerable advantage since a direct calculation of the classical action (by integrating the classical Lagrangian) is in general a tedious job.
In the presence of a magnetic field $\bold{B}$, derivable from a vector potential $\bold{A}=\bold{\nabla}\times \bold{B}$,  the propagator becomes\cite{schulman:1981} ($\bold{r}=(x,y)$):
\begin{equation}
\label{propagator}
K(\bold{r_f},T;\bold{r_i},0)=\mathcal{N}\int \mathcal{D}[\bold{r}(t)] \mathrm{e}^{\frac{i}{\hbar}\int\limits_0^T dt[\frac{m}{2}\dot{\bold{r}}^2-V(\bold{r},t)+q\dot{\bold{r}}\bold{A(r)}]}
\end{equation}
The translational invariant measure $\mathcal{D}[\bold{r}(t)]$ in Eq.~(\ref{propagator}) suggests an integration over the set of all continuous paths $\bold{r}(t)$ starting in $\bold{r}(0)=\bold{r_i}$ and ending in $\bold{r}(T)=\bold{r_f}$.\\The standard procedure\cite{feynman:48} of calculating this integral is by dividing up the time interval $[0,T]$ in N time slices. In the limit $N\rightarrow \infty$, the integral over positions at each time slice can be said to be an integral over all possible continous paths. 
However, the method has from a mathematical perspective several conceptional drawbacks: The continous paths over which one integrates in  Eq.~(\ref{propagator}) are almost certainly nowhere differentiable. This means that kinetic energy is a diverging quantity at all times. 
An additional drawback is the occurrence of an infinite Normalization constant $\mathcal{N}$ in the limit $N\rightarrow \infty$  of infinitely many time slices.\\
Evaluation of path integrals for Lagrangians with a velocity dependent potential, such as the magnetic interaction by the method of time discretizing is more delicate; it leads to ambiguities when not correctly discretized. A correct calculation requires 
the vector potential $\bold{A(r)}$ be evaluated at the midpoint of each straight segment of the path. (See Ref.\cite{schulman:1981}). However, by applying either one of the above described methods one can avoid such difficulties.
This paper is organized in the following manner: In Sec.\eqref{eval_path_complex} the method of path integral calculation by complex Fourier series is presented (\emph{$1^{\text{st}}$} method) whereas in Sec.\eqref{eval_rot} the alternative rotational method is illustrated (\emph{$2^{\text{nd}}$} method). Both methods appear to be novel. In Sec.\eqref{spectrum} the energy spectrum in the isotropic case is calculated. 
\section{\label{lagrange}Lagrangian}
By choosing the symmetric gauge for the vector potential $\bold{A(r)}$ the Lagrangian $L$ becomes:
\begin{equation}
\label{general_lagra}
L={\frac{m}{2}(\dot{x}^2+\dot{y}^2)}-{\frac{m\omega_{x}^2}{2} x^2-\frac{m\omega_{y}^2}{2} y^2}
+m\omega_{L}(x\dot{y}-y\dot{x})+q\bold{E}\cdot\bold{r}
\end{equation}
where $\omega_{L}=\frac{qB}{2m}$ is the Larmor frequency and $\omega_{x},\omega_{y}$ are the oscillator frequencies in each direction of the plane.
Eq.~(\ref{general_lagra}) describes the Lagrangian of a charged anisotropic Harmonic Oscillator (HO) submitted to both a homogeneous magnetic field aligned with the perpendicular direction to the plane and a static electric field within the plane. The corresponding propagator is calculated in full generality in Sec.\eqref{ho_electric_magnetic}.\\
In case of an isotropic harmonic potential $\omega_x=\omega_y$ it is convenient to express Eq.~(\ref{general_lagra}) in a complex form rather than in Cartesian "coordinates":
\begin{equation}
\gamma(t):=x(t)+iy(t)\quad \text{and}\quad \epsilon:=E_x+iE_y
\end{equation}
where $\gamma(t)$ is a complex-valued function defined on the time interval $[0,T]$ which describes a path in the complex plane. $\gamma(t)$ is chosen from the Hilbert space $L^2_{\mathbb{C}}[0,T]$, (\ie the space of square-integrable complex-valued functions on $[0,T]$). In terms of $\gamma(t)$ the Lagrangian Eq.~(\ref{general_lagra}) is of the following form:
\begin{equation}
\label{complex_lagra}
L=\frac{m}{2}\dot{\gamma}\dot{\overline{\gamma}}-\frac{m \omega^2}{2}\gamma\overline{\gamma} +\frac{im\omega_{L}}{2}(\gamma\dot{\overline{\gamma}}-\dot{\gamma}\overline{\gamma})+\frac{q}{2}\epsilon\overline{{\gamma}}+\frac{q}{2}\overline{\epsilon}\gamma
\end{equation}
The action $S$ can be expressed as a sum of inner products on $L^2_{\mathbb{C}}[0,T]$ 
\begin{equation}
\label{action}
S[\gamma]=T\frac{m}{2}\langle \dot{\gamma}|\dot{\gamma}
\rangle-\frac{m}{2}\omega^2 T \langle\gamma|\gamma\rangle+\frac{im\omega_{L}T}{2}[\langle\gamma|\dot{\gamma}\rangle
-\langle\dot{\gamma}|\gamma\rangle]
\end{equation}
where the inner product $\langle|\rangle$ is defined in the usual form:
\begin{equation}
\langle f|g\rangle:=\frac{1}{T}\int\limits_0^T f(t)\cdot \overline{g(t)}dt
\end{equation}
The propagator then reads:
\begin{equation}
K({b},T;{a},0)=\int\limits_{\gamma(0)={a}}^{\gamma(T)={b}}\mathcal{D}[\gamma(t)]\exp\left[{\frac{i}{\hbar}S[\gamma]}\right]
\end{equation}
where boundary conditions are given by:
\begin{equation}
\label{boundary}
\begin{split}
 \gamma(0)&={a}\equiv x_{a}+ib_{a}\in \mathbb{C}\\
 \gamma(T)&={b}\equiv x_{b}+iy_{b}\in \mathbb{C}\\
\end{split}
\end{equation}
$\mathcal{D}[\gamma(t)]$ denotes the translational and rotational invariant measure defined (heuristically) in the next section by using an infinite product of complex Fourier coefficients.
\section{\label{eval_path_complex}Evaluation by complex Fourier Series}
\subsection{\label{ho}The Harmonic Oscillator}
The present paper is didactic in spirit. In order to introduce the calculational method the author evaluates therefore the simplest non-trivial case of an isotropic Harmonic Oscillator. The propagator reads here:
\begin{equation}
\label{general_propagator}
K({b},T;{a},0)= 
\int\limits_{\gamma(0)={a}}^{\gamma(T)={b}}\mathcal{D}[\gamma(t)]
\exp\left[\frac{imT}{2\hbar}\langle \dot{\gamma}|\dot{\gamma}
\rangle-\frac{im\omega^2 T}{2\hbar}\langle\gamma|\gamma\rangle\right]
\end{equation}
In the following we split any path satisfying the boundary conditions Eqs.~(\ref{boundary}) into the sum of the classical path $\gamma_{\text{c}}(t)$ and a \emph{quantum} fluctuation $\delta \gamma(t)$:
\begin{equation}
\label{variable_trans}
\gamma(t) = \gamma_{\text{c}}(t)+\delta \gamma(t)
\end{equation}
with $\delta \gamma(t)$ satisfying the boundary (Dirichlet) conditions:
\begin{equation}
\delta \gamma(0)=\delta \gamma(T)=0
\end{equation}
The function $\delta \gamma(t)$ describes a closed path (loop) in the complex plane. The transformation Eq.~(\ref{variable_trans}) corresponds to an overall shift by a constant "amount" $\gamma_{\text{c}}$. The Jacobian of this change of variables is therefore assumed to be 1. From the translational invariance of the measure $\mathcal{D}[\gamma(t)]=\mathcal{D}[\delta \gamma(t)] $ one obtains:
\begin{equation}
\label{prop_shifted}
\begin{split}
K&= \exp\left[\frac{i}{\hbar}S_{\text{c}}\right]
\int\limits_{\delta \gamma(0)=0}^{\delta \gamma(T)=0}\mathcal{D}[\delta \gamma(t)]
\exp\left[\frac{imT}{2\hbar}\langle \dot{\delta \gamma}|\dot{\delta \gamma}
\rangle-\frac{im\omega^2 T}{2\hbar}\langle\delta \gamma|\delta \gamma\rangle\right]\\
&\equiv \exp\left[\frac{i}{\hbar}S_{\text{c}}\right] F(T)
\end{split}
\end{equation}
Where $F(T)(\equiv K(0,T;0,0))$ denotes the \emph{fluctuation} integral of the free particle confined to the plane. In the following one solves the integral Eq.~(\ref{prop_shifted}) by expanding the fluctuation $\delta \gamma(t)$ in a complex Fourier series. This is indeed possible since $\delta \gamma(t)$ describes a loop in the complex plane. It is therefore possible to continue $\delta \gamma(t)$ to a T-periodic (complex-valued) function on $\mathbb{R}$. The author borrowed the idea of Fourier expanding a function describing a closed path in the plane from Hurwitz' elegant solution of the isoperimetric problem.\cite{hurwitz:1902}(See also Ref.\cite{courant:43})\\
The complex Fourier expansion reads then:
\begin{equation}
\label{fourier_exp}
\delta \gamma(t)=\sum_{n=-\infty}^{+\infty}c_{n}\exp\left[i\frac{2\pi}{T}n\cdot t\right], \quad c_n
\in \mathbb{C}
\end{equation} 
On the other hand, the action S is given by:
\begin{equation}
S[\delta \gamma,\dot{\delta \gamma}]=\frac{mT}{2}\langle \dot{\delta \gamma}|\dot{\delta \gamma}\rangle-\frac{m\omega^2 T}{2}\langle\delta \gamma|\delta \gamma\rangle
\end{equation}
Plugging Eq.~(\ref{fourier_exp}) into the action S one finds (Parseval's identity):
\begin{subequations}
\begin{equation}
\langle\delta \gamma|\delta \gamma\rangle=\sum_{n=-\infty}^{+\infty}|c_{n}|^2,
\end{equation}
and
\begin{equation}
\label{parseval2}
\langle \dot{\delta \gamma}|\dot{\delta \gamma}\rangle= \left(\frac{2\pi}{T}\right)^2 \sum_{n=-\infty}^{+\infty}|c_{n}|^2 n^2 
\end{equation}
\end{subequations}
The latter series is in general divergent. According to the theorem of Riesz-Fischer (See \eg Ref.\cite{werner:2011}) one can only guarantee the following:  
\begin{equation}
\label{Riesz-Fischer}
\delta \gamma(t) \in L^2_{\mathbb{C}}[0,T] \Longleftrightarrow \sum_{n=-\infty}^{+\infty}|c_{n}|^2 < \infty
\end{equation}
For the moment one can look at Eq.~(\ref{parseval2}) as a purely formal expression. A possibility to cure this mathematical ill is illustrated in App. \eqref{remark_convergence}.\\
Thus, on formal grounds one writes:\\
\begin{equation}
\label{fluctuation_ho}
F(T)=\int\prod_{n=-\infty}^{+\infty}dc_{n}
\exp{\left[i\sum_{n=-\infty}^{+\infty}|c_{n}|^2 \left(\alpha n^2-\beta\right)\right]}
\delta\left(\sum\limits_{n=-\infty}^{+\infty}c_n\right)\\
\end{equation}
where the parameters are given by:
\begin{equation}
\alpha:=\frac{2m\pi^2}{\hbar T} \quad \beta:=\frac{m\omega^2 T}{2\hbar}
\end{equation}
The Dirac-Heaviside $\delta$-Function\cite{dirac:1930} has been introduced in order to fullfill the boundary conditions:
\begin{equation}
\label{boundary_eta}
\delta \gamma(0)=\sum\limits_{n=-\infty}^{+\infty}c_n=0=\delta \gamma(T)
\end{equation}
Eq.~(\ref{fluctuation_ho}) is an infinite-dimensional divergent Fresnel integral which is only meaningful when regularized in a suitable way. Somewhat unexpectedly, a regularization via the $\zeta$-function leads straightforwardly to the desired results. For that purpose one needs the values of $\zeta(0)$ and $\zeta^{\prime}(0)$ given by:
\begin{equation}
\label{zeta_values}
\zeta(0)=-\frac{1}{2} \quad \text{and} \quad 
\zeta^{\prime}(0)=-\frac{1}{2}\log{2\pi}
\end{equation}
for an heuristic derivation of these values see App.\eqref{values_zeta}. Moreover, the complex integration in Eq.~(\ref{fluctuation_ho}) is defined in the following canonical way:
\begin{equation} 
\int_{\mathbb{C}} dc_{n}:=\int_{\mathbb{R}^2} dx_{n} dy_{n}
\end{equation}
By using the identity:
\begin{equation}
\delta\left(\sum\limits_{n=-\infty}^{+\infty}c_n\right)=\delta\left(\sum\limits_{n=-\infty}^{+\infty}x_n\right)\delta\left(\sum\limits_{n=-\infty}^{+\infty}y_n\right)
\end{equation}
the integration over $x_n$, $y_n$ in Eq.~(\ref{fluctuation_ho}) can be calculated separately so that one has:
\begin{equation}
F(T)=[F_{1\text{d}}]^2
\end{equation}
Where $F_{1\text{d}}$ describes the \emph{fluctuation} integral in one dimension. Explicitly:
\begin{equation}
\label{fluctuation_1d}
F_{1\text{d}}(T)=\int\prod_{n=-\infty}^{+\infty}dx_{n}
\exp{\left[i \sum_{n=-\infty}^{+\infty}|x_{n}|^2 \left(\alpha n^2-\beta\right)\right]}
\delta\left(\sum\limits_{n=-\infty}^{+\infty}x_n\right)
\end{equation}
By inserting the Integral representation of $\delta$-Function and by assuming somewhat sloppily that the interchange of $x_n$ and $k$-integration poses no problem, one obtains: 
\begin{equation}
\label{calculation}
\begin{split} 
F_{1\text{d}}(T)&=\frac{1}{2\pi}\int dk \int\prod dx_{n}\exp \left[i \sum x_{n}^2 (\alpha n^2-\beta)+kx_{n}\right]\\
&= \frac{1}{2\pi}\int dk \prod_{n=-\infty}^{+\infty} \underbrace{\int dx_{n}\exp \left[i \sum x_{n}^2 (\alpha n^2-\beta)+kx_{n}\right]}_{\text{Fresnel Integral}}\\
&= \frac{1}{2\pi}\int dk \prod_{n=-\infty}^{+\infty} \exp\left(\frac{-ik^2}{4(\alpha n^2-\beta)}\right)\sqrt{\frac{i\pi}{\alpha n^2-\beta}}\\
&= \frac{1}{2\pi}\int dk \underbrace{\prod_{n=-\infty}^{+\infty} \exp\left(\frac{-ik^2}{4(\alpha n^2-\beta)}\right)}_{\triangle}\underbrace{\prod_{n=-\infty}^{+\infty}\sqrt{\frac{i\pi}{\alpha n^2-\beta}}}_{\bigstar}
\end{split} 
\end{equation}
Both products in Eq.~(\ref{calculation}) can be evaluated in a straigtforward manner.\cite{gradshteyn:2014} On one hand:
\begin{equation}
\label{triangle}
\begin{split}
\log \triangle&=\sum\limits_{n=-\infty}^{+\infty}\frac{-i k^2}{4(\alpha n^2-\beta)}
=\frac{-i k^2}{4\alpha}\sum\limits_{n=-\infty}^{+\infty}\frac{1}{n^2-\frac{\beta}{\alpha}}\\
&=\frac{i k^2 \pi}{4\sqrt{\alpha \beta}} \cot{\pi \sqrt{\frac{\beta}{\alpha}}}
\end{split}
\end{equation}
On the other hand:
\begin{equation}
\label{star}
\begin{split}
\bigstar&=\prod_{n=-\infty}^{+\infty}\sqrt{\frac{i\pi}{\alpha n^2-\beta}}
=\sqrt{\frac{i \pi}{-\beta}}\prod_{n=1}^{+\infty}\frac{i\pi}{\alpha n^2-\beta}\\
&=\sqrt{\frac{i \pi}{-\beta}}\prod_{n=1}^{+\infty}\frac{i\pi}{\alpha n^2}\underbrace{\prod_{n=1}^{+\infty}\frac{1}{1-\frac{\beta}{\alpha n^2}}}_{\text{Eulerian Sine product}}
\end{split}
\end{equation}
The latter product is the well-known Eulerian sine product\cite{ahlfors:78} whereas the former product has to be $\zeta$-regularized. By taking the Logarithm of the product one obtains formally:
\begin{equation}
\log{\left(\prod_{n=1}^{+\infty}\frac{i\pi}{\alpha n^2}\right)}=\log{\left(\frac{i\pi}{\alpha}\right)}\sum_{n=1}^{+\infty}1-2\sum_{n=1}^{+\infty}\log{n}
\end{equation}
By considering the following $\zeta$-regularized identities:
\begin{equation}
\begin{split}
&\sum_{n=1}^{+\infty}\log{n}=\lim_{x\to 0}\left(-\frac{d}{dx}\sum_{n=1}^{+\infty}\exp{\left[-x\log{n}\right]}\right)=-\zeta^{\prime}(0)\\
&\sum_{n=1}^{+\infty}1=\zeta(0)
\end{split}
\end{equation}
one arrives by Eq.~(\ref{zeta_values}) immediately at:
\begin{equation}
\prod_{n=1}^{+\infty}\frac{i\pi}{\alpha n^2}
=\frac{1}{2\pi}\sqrt{\frac{\alpha}{i \pi}}
\end{equation}
By substituting the derived expressions for $\triangle$ and $\bigstar$  into Eq.~(\ref{calculation}) one gets the expected result:
\begin{equation}
\label{fluctuation_result} 
F_{1\text{d}}(T)=\frac{1}{2\pi}\sqrt{\frac{2\alpha^{\frac{1}{2}}\beta^{\frac{1}{2}}}{i \sin{(2\pi\sqrt{\frac{\beta}{\alpha}})}}}=\sqrt{\frac{m \omega}{2 \pi i \hbar \sin{\omega T}}}
\end{equation}
As is well-known the \emph{classical} action $S_{\text{c}}$ for the 1D-HO reads (See \eg Ref.\cite{feynman:2010}):
\begin{equation}
S_{\text{c}}=\frac{m\omega}{2\sin{\omega T}}\left[(x_a^2+x_b^2)\cos{\omega T}-2x_a x_b\right]
\end{equation}
In the limit $\omega\rightarrow 0$ one recovers the propagator of the free particle:
\begin{equation}
K_{0}(x_b,T;x_a,0)= \sqrt{\frac{m}{2\pi i\hbar T}}\exp{\left[\frac{im\left(x_b-x_a\right)^2}{2\hbar T}\right]}
\end{equation}
\subsection{Charged Harmonic Oscillator in a constant magnetic field}
\label{3B}
In this section we consider the more general case of a charged particle submitted to both a harmonic oscillator in the plane and a constant and homogeneous magnetic field in the perpendicular direction.
The evaluation of the path integral for Lagrangians with a velocity dependent potential, such as the magnetic interaction is usually done by discretizing the time interval. As already described in the introduction, this method leads to ambiguities when not correctly discretized.
A direct evaluation of magnetic-field including path integrals by Fourier expansion is to the author's knowledge absent in the literature. One sees in App. \eqref{failure_real_fourier} that an evaluation by a Fourier series based on the eigenmodes Eq.~(\ref{real_expansion})        
(including a $\zeta$-Regularization) leads indeed to incorrect results.\\
In this section the author shows that an evaluation by Fourier Series is possible and rather convenient, provided one uses the complex form given by Eq.~(\ref{fourier_exp}) instead of Eq.~(\ref{real_expansion}). With the definitions introduced in previous Sec.\eqref{ho} this is a simple and rather short calculation.\\
The \emph{fluctuation} integral reads in this case:
\begin{equation}
\label{fluctuation_charged_ho}
\begin{split}
F(T)&=\int\prod_{n=-\infty}^{+\infty}dc_{n}
\exp{\left[i\sum_{n=-\infty}^{+\infty}|c_{n}|^2 (\alpha n^2-\beta+\gamma n)\right]}
\delta\left(\sum\limits_{n=-\infty}^{+\infty}c_n\right)
\end{split} 
\end{equation}
where the parameters are given by:
\begin{equation}
\alpha:=\frac{2m\pi^2}{\hbar T} \quad \beta:=\frac{m\omega^2 T}{2\hbar}\quad \gamma:=\frac{2m\pi\omega_{L}}{\hbar} 
\end{equation}
Once again, the integral Eq.~(\ref{fluctuation_charged_ho}) splits separately into an integration over $x_{n}$ and $y_{n}$. In analogy to Eq.~(\ref{calculation}) one obtains:
\begin{equation}
\label{root_fluctuation}
\begin{split} 
\sqrt{F(T)}&=\int\prod_{n=-\infty}^{+\infty}dx_{n}
\exp{\left[i\sum_{n=-\infty}^{+\infty}x^2_{n} (\alpha n^2-\beta+\gamma n)\right]}
\delta\left(\sum\limits_{n=-\infty}^{+\infty}x_n\right)\\
&=\frac{1}{2\pi}\int dk \prod_{\substack{n=-\infty}}^{+\infty} \exp\left[\frac{-ik^2}{4(\alpha n^2-\beta+\gamma n)}\right]\sqrt{\frac{i\pi}{\alpha n^2-\beta+\gamma n}}\\
&=\frac{1}{2\pi}\int dk \underbrace{\prod_{\substack{n=-\infty}}^{+\infty} \exp\left[\frac{-ik^2}{4(\alpha n^2-\beta+\gamma n)}\right]}_{\triangle\triangle}\underbrace{\prod_{\substack{n=-\infty}}^{+\infty}\sqrt{\frac{i\pi}{\alpha n^2-\beta+\gamma n}}}_{\bigstar\bigstar}
\end{split} 
\end{equation}
Once again, both products are easily evaluated (See Ref.\cite{gradshteyn:2014}):
\begin{equation}
\begin{split}
\log \triangle\triangle&=-\frac{i k^2}{4}\sum\limits_{n=-\infty}^{+\infty}\frac{1}{\alpha n^2-\beta+\gamma n}\\
&=\frac{i\pi k^2}{4\sqrt{\gamma^2+4\alpha \beta}}\left[\cot{\pi a_{+}}-\cot{\pi a_{-}}\right]
\end{split}
\end{equation}
where:
\begin{equation}
\begin{split}
\pi a_{+}&=\frac{1}{2}\left(\sqrt{\omega^2+\omega_{L}^2}-\omega_{L}\right)T\equiv \omega_{\text{eff}}^{-} T\\
\pi a_{-}&=-\frac{1}{2}\left(\sqrt{\omega^2+\omega_{L}^2}+\omega_{L}\right)T\equiv -\omega_{\text{eff}}^{+} T\\
\end{split}
\end{equation}
Integration over the k-variable in Eq.~(\ref{root_fluctuation}) yields:\\
\begin{equation}
\label{doubletriangle}
\begin{split}
\int_{\mathbb{R}}\triangle\triangle dk &=
\int_{\mathbb{R}} dk \exp\left[-\frac{i k^2}{4}\sum\limits_{n=-\infty}^{+\infty}\frac{1}{\alpha n^2-\beta+\gamma n}\right]\\
&=\sqrt{\frac{4i \sqrt{\gamma^2+4\alpha \beta}}{\cot{\omega_{\text{eff}}^+ T}+\cot{\omega_{\text{eff}}^- T}}}
\end{split}
\end{equation}\\
Once again, by $\zeta-$regularizating the product $\bigstar\bigstar$ one gets:
\begin{equation}
\label{doublestar}
\bigstar\bigstar={\prod_{\substack{n=-\infty}}^{+\infty}\sqrt{\frac{i\pi}{\alpha n^2-\beta+\gamma n}}}=\sqrt{\frac{imT\omega_{\text{eff}}^+\omega_{\text{eff}}^-}{2\pi\hbar\sin{\omega_{\text{eff}}^+T}\sin{\omega_{\text{eff}}^-T}}}
\end{equation}\\
By multiplying \eqref{doubletriangle} with \eqref{doublestar} and squaring the result one obtains the remarkable simple result:
\begin{equation}
\label{fluctuation_ho_b}
F(T)=\frac{m\sqrt{\omega^2+\omega_{L}^2}}{2\pi i\hbar\sin\left(T\sqrt{\omega^2+\omega_{L}^2}\right)}\equiv 
\frac{m\omega_{\text{eff}}}{2\pi i\hbar\sin{\omega_{\text{eff}}T}}
\end{equation}
The simple formula \eqref{fluctuation_ho_b} does not seem to occur in the given symmetric form in standard literature.\cite{feynman:2010, schulman:1981, kleinert:2004}
The corresponding \emph{classical} action $S_{\text{c}}$ will be calculated in the following Sec.\eqref{eval_rot_charged_ho} in a rather simple manner by a different method. 
\section{Evaluation by Rotation and Translation}
\label{eval_rot}
\subsection{Charged HO in a constant magnetic field}
\label{eval_rot_charged_ho}
In this section the author presents a remarkably simple technique based on the fact that the Lorentz-and Coriolisforce cancel each other in a suitably chosen rotating system. By applying this method one can transform a Lagrangian with a velocity dependent term such as the magnetic field to a Lagrangian of an effective harmonic oscillator.\\
As described above the Lagrangian reads:
\begin{equation}
L=\frac{m}{2}\dot{\gamma}\dot{\overline{\gamma}}+\frac{im\omega_{L}}{2}(\gamma\dot{\overline{\gamma}}-\dot{\gamma}\overline{\gamma})-\frac{m}{2}\omega^2\gamma\overline{\gamma}
\end{equation}
The uniform rotation with frequency $\Omega$ is described by:
\begin{equation}
\gamma(t)=\exp\left(i \Omega t\right) \gamma_{\text{rot}}(t)
\end{equation}
The transformed Lagrangian thus reads:
\begin{equation}
\begin{split}
L_{\text{rot}}&=\frac{m}{2}\dot{\gamma}_{\text{rot}}\dot{\overline{\gamma}}_{\text{rot}}+ \frac{m}{2}(\Omega^2+2\Omega\omega_{L}) \gamma_{\text{rot}}\bar{\gamma}_{\text{rot}}
-\frac{m}{2}\omega^2\gamma_{\text{rot}}\overline{\gamma}_{\text{rot}}\\
&+\frac{im(\Omega+\omega_{L})}{2}(\gamma_{\text{rot}}\dot{\overline{\gamma}}_{\text{rot}}-\dot{\gamma}_{\text{rot}}\overline{\gamma}_{\text{rot}})
\end{split}
\end{equation}
By fixing $\Omega=-\omega_{L}$ the Lagrangian assumes the form of an effective HO:
\begin{equation}
\label{lagrange_rot}
L_{\text{rot}}=\frac{m}{2}\dot{\gamma}_{\text{rot}}\dot{\overline{\gamma}}_{\text{rot}}-\frac{m\omega_{\text{eff}}^2}{2}  \gamma_{\text{rot}}\bar{\gamma}_{\text{rot}}
\end{equation}
where $\omega_{\text{eff}}:=\sqrt{\omega^2+\omega_{L}^2}$\\ 
The suitable transformation is therefore given by:
\begin{equation}
\gamma_{\text{rot}}(t)=\exp\left(i \omega_{L} t\right) \gamma(t)
\end{equation}
describing a system rotating with Larmorfrequency $\omega_{L}=qB/2m$ in the counter-clockwise direction.
Presuming rotational invariance of the functional measure:
\begin{equation}
\mathcal{D}[\gamma_{\text{rot}}(t)]=\mathcal{D}[\gamma(t)] 
\end{equation}
one gets straightforwardly:
\begin{equation}
\label{prop_rot}
\begin{split}
K(b,T;a,0)&=\int\limits_{\gamma(0)}^{\gamma(T)} \mathcal{D}[\gamma(t)] \exp{\left[\frac{i}{\hbar}{S}\right]} =\int\limits_{\gamma_{\text{rot}}(0)}^{\gamma_{\text{rot}}(T)} \mathcal{D}[{\gamma_{\text{rot}}}(t)]\exp{\left[\frac{i}{\hbar}{{S_{\text{rot}}}}\right]}\\
&=\exp{\left[\frac{i}{\hbar}{{\bar{S}}_{\text{c}}}\right]} \frac{m\omega_{\text{\text{eff}}}}{2\pi i\hbar \sin{\omega_{\text{\text{eff}}}T}}
\end{split}
\end{equation}
where $\bar{S}_{\text{c}}$ denotes the \emph{classical} action of a two-dimensional effective HO within the rotating frame. Therefore, 
in terms of transformed variables $\bar{x}$, $\bar{y}$ one concludes immediately:  
\begin{equation}
\label{classical_action}
\begin{split}
{\bar{S}}_{\text{c}}&=\frac{m\omega_{\text{eff}}}{2\sin{{\omega_{\text{eff}}T}}} \left[(\bar{x}^2_{a}+\bar{x}^2_{b}+\bar{y}^2_{a}+\bar{y}^2_{b})\cos{\omega_{\text{eff}}T}-2\bar{x}_{a}\bar{x}_{b}-2\bar{y}_{a}\bar{y}_{b}\right]\\
\end{split}
\end{equation}
The boundary conditions in terms of untransformed variables are given by:
\begin{equation}
\label{variable_transform}
\begin{split}
\bar{x}_a&=x_{a}, \quad
\bar{y}_{a}= y_{a}\\
\bar{x}_{b}&=x_{b}\cos{\omega_{L}T} -y_{b}\sin{\omega_{L}T}\\                                            
\bar{y}_{b}&=x_{b}\sin{\omega_{L}T} +y_{b} \cos{\omega_{L}T}                                          
\end{split}
\end{equation}\\
Plugging Eqs.~(\ref{variable_transform}) into Eq.~(\ref{classical_action}) one gets the \emph{classical} action $S_{\text{c}}$: 
\begin{equation}
\label{classical_action_ho_b}
\begin{split}
{S}_{\text{c}}&=\frac{m\omega_{\text{\text{eff}}}}{2\sin{{\omega_{\text{\text{eff}}}T}}}[\left({x}^2_{a}+{x}^2_{b}+{y}^2_{a}+{y}^2_{b}\right)\cos{\omega_{\text{\text{eff}}}T}\\&-2(x_a x_b+y_a y_b)\cos{\omega_{L}T}-2(y_a x_b-x_a y_b)\sin{\omega_{L}T}]
\end{split}
\end{equation}
The full propagator is therefore given by:
\begin{equation}
K=\frac{m\sqrt{\omega^2+\omega_{L}^2}}{2\pi i\hbar \sin(T\sqrt{\omega^2+\omega_{L}^2})}\exp\left[{\frac{i}{\hbar}{S}_{\text{c}}}\right]
\end{equation}
In the limit $\omega\rightarrow0$ the \emph{classical} action Eq.~(\ref{classical_action_ho_b}) reduces to the well-known formula\cite{feynman:2010,schulman:1981}:
\begin{equation}
{S}_{\text{c}}^{\text{B}}=\frac{m\omega_{L}}{2\tan{{\omega_{L}T}}}\left[(x_b-x_a)^2 + (y_b-y_a)^2\right]-m\omega_{L}(y_a x_b-x_a y_b)
\end{equation}
\subsection{charged anisotropic HO in crossed electric and magnetic fields}
\label{ho_electric_magnetic}
In this section the most general case of this paper is calculated: The HO-Potential is allowed to be anisotropic. In addition, one assumes a constant electric field in the plane. As already stated, the considered Lagrangian is of the form:
\begin{equation}
L={\frac{m}{2}(\dot{x}^2+\dot{y}^2)}-{\frac{m\omega_{x}^2 }{2}x^2-\frac{m\omega_{y}^2 }{2}y^2}
+m\omega_{L}(x\dot{y}-y\dot{x})+q\bold{E}\cdot\bold{r}
\end{equation}
By completing the square one arrives at:
\begin{equation}
\begin{split}
L&={\frac{m}{2}(\dot{x}^2+\dot{y}^2)}-\frac{m\omega_{x}^2 }{2}\left[x-\frac{qE_x}{m\omega_{x}^2}\right]^2
-\frac{m\omega_{y}^2 }{2}\left[y-\frac{qE_y}{m\omega_{y}^2}\right]^2\\
&+m\omega_{L}\left(x\dot{y}-y\dot{x}\right)+\frac{q^2E_{x}^2}{2m\omega_{x}^2}+\frac{q^2E_{y}^2}{2m\omega_{y}^2}
\end{split}
\end{equation}
By the following variable shift:
\begin{equation}
\begin{split}
\bar{x}=x-\frac{qE_x}{m\omega_x^2}\\
\bar{y}=y-\frac{qE_y}{m\omega_y^2}
\end{split}\
\end{equation}
one obtains the Lagrangian $\bar{L}$ in terms of shifted variables $\bar{x}, \bar{y}$:
\begin{equation}
\begin{split}
\bar{L}&={\frac{m}{2}(\dot{\bar{x}}^2+\dot{\bar{y}}^2)}-\frac{m\omega_{x}^2 }{2}\bar{x}^2
-\frac{m\omega_{y}^2 }{2}\bar{x}^2
+m\omega_{L}\left(\bar{x}\dot{\bar{y}}-\bar{y}\dot{\bar{x}}\right)\\
&+\frac{q^2E_{x}^2}{2m\omega_{x}^2}+\frac{q^2E_{y}^2}{2m\omega_{y}^2}
+\frac{q\omega_L E_x}{\omega_x^2}\dot{\bar{y}}-\frac{q\omega_L E_y}{\omega_y^2}\dot{\bar{x}}
\end{split}
\end{equation}
The action splits naturally in two parts:
\begin{equation}
\begin{split}
\bar{S}=\int \bar{L} dt&=\frac{q^2 T}{2m}\left(\frac{E_x^2}{\omega_x^2}+\frac{E_y^2}{\omega_y^2}\right)
+\frac{q\omega_\text{L}E_x}{\omega_x^2}\left(y_b-y_a\right)-\frac{q\omega_\text{L}E_y}{\omega_y^2}\left(x_b-x_a\right)\\
&+\int \bar{L_t} dt
\equiv C+\bar{S}_t
\end{split}
\end{equation}
where: 
\begin{equation}
\bar{L_t}={\frac{m}{2}(\dot{\bar{x}}^2+\dot{\bar{y}}^2)}-\frac{m\omega_{x}^2 }{2}\bar{x}^2
-\frac{m\omega_{y}^2 }{2}\bar{x}^2
\end{equation}
By presuming (as usual) translational invariance of the functional measure:
\begin{equation}
\mathcal{D}[\bar{x}(t)]\mathcal{D}[\bar{y}(t)]=\mathcal{D}[x(t)]\mathcal{D}[y(t)] 
\end{equation}
one gets readily:
\begin{eqnarray}
\label{prop_electric_magnetic}
K=\int\mathcal{D}[x(t)]\mathcal{D}[y(t)]\exp\left[{\frac{i}{\hbar}S}\right]
=\exp{\left[\frac{i}{\hbar}C\right]}\int\mathcal{D}[\bar{x}(t)]\mathcal{D}[\bar{y}(t)]\exp\left[{\frac{i}{\hbar}\bar{S}_t[\gamma]}\right]
\end{eqnarray}
In Eq.~(\ref{prop_electric_magnetic}) we are faced with a Lagrangian where the electric field is effectively absent. 
By the following continous rotation (with Larmor frequency $\omega_\text{L}$): 
 \[
    \begin{pmatrix} \bar{x} \\ \bar{y} \end{pmatrix} =
    \begin{pmatrix} \cos \omega_{L}t& \sin\omega_{L}t \\
        -\sin\omega_{L}t & \cos\omega_{L}t \end{pmatrix}\begin{pmatrix} \tilde{x} \\ \tilde{y} \end{pmatrix}
  \]
one gets an even more simplified form:
\begin{equation}
{\tilde{L}}=\frac{m}{2}(\dot{\tilde{x}}^2+\dot{\tilde{y}}^2)-  \frac{m \omega^{x}_{\text{eff}}}{2} \tilde{x}^2
     -  \frac{m \omega^{y}_{{\text{eff}}}}{2} {\tilde{y}}^2
\end{equation}
which is a Lagrangian of an anisotropic HO in zero magnetic and electric field with effective frequencies given by:
\begin{equation}
\omega^{x}_{\text{eff}}=\sqrt{\omega_{L}^2 +\omega_{x}^2} \quad \omega^y_{\text{eff}}=\sqrt{\omega_{L}^2 +\omega_{y}^2}
\end{equation}
Since the propagator of the anisotropic HO is well-known one concludes immediately:
\begin{equation}
\label{prop_anis_ho}
\begin{split}
K&=\exp{\left[\frac{i}{\hbar}(C+\tilde{S}_{\text{c}}^{\tilde{x}}+\tilde{S}_{\text{c}}^{\tilde{y}})\right]}\frac{m}{2\pi i \hbar}
\sqrt{\frac{\omega^{x}_{\text{eff}}\omega^{y}_{\text{eff}}}{\sin{\omega^{x}_{\text{eff}}T}\sin{\omega^{y}_{\text{eff}}T}}}
\end{split}
\end{equation}
where:
\begin{equation}
\label{prop_electric_magnetic_final}
\begin{split}
\tilde{S}_{\text{c}}^{\tilde{x}}&=\frac{m\omega^{{x}}_{\text{eff}}}{2\sin{{\omega^{{x}}_{\text{eff}}T}}} \left[(\tilde{x}^2_{a}+\tilde{x}^2_{b})\cos{\omega^{{x}}_{\text{eff}}T}-2\tilde{x}_{a}\tilde{x}_{b}\right]\\
\tilde{S}_{\text{c}}^{\tilde{y}}&=({x} \Leftrightarrow {y})
\end{split}
\end{equation}
In order to express the \emph{classical} action in terms of untransformed variables $x,y$ one should replace the variables $\tilde{x}, \tilde{y}$ in Eq.~(\ref{prop_electric_magnetic_final}) by:
\begin{equation}
\begin{split}
\tilde{x}_a&=x_a-\frac{qE_x}{m\omega_x^2} \quad \tilde{y}_a=y_a-\frac{qE_y}{m\omega_y^2}\\
\tilde{x}_b&=\left(x_b-\frac{qE_{x}}{m\omega_x^2}\right)\cos{\omega_{L}T}-\left(y_b-\frac{qE_{y}}{m\omega_y^2}\right)\sin{\omega_{L}T}\\
\tilde{y}_b&=\left(x_b-\frac{qE_{x}}{m\omega_x^2}\right)\sin{\omega_{L}T}+\left(y_b-\frac{qE_{y}}{m\omega_y^2}\right)\cos{\omega_{L}T}
\end{split}
\end{equation}
A general calculation is omitted here, restricting to limiting cases in the following analysis.
\subsubsection{Vanishing electric field $\bold{E}=0$:}
A short calculation yields
\begin{equation}
\label{prop_anisotrop_ho}
\begin{split}
K&=\exp{\left[\frac{i}{\hbar}{{S}_{\text{c}}^{{x}}}\right]}\exp{\left[\frac{i}{\hbar}{{S}_{\text{c}}^{{y}}}\right]}\frac{m}{2\pi i \hbar}
\sqrt{\frac{\omega^{x}_{\text{eff}}\omega^{y}_{\text{eff}}}{\sin{\omega^{x}_{\text{eff}}T}\sin{\omega^{y}_{\text{eff}}T}}}
\end{split}
\end{equation}
where the \emph{classical} action reads: 
\begin{subequations}
\begin{equation}
\begin{split}
S_{\text{c}}^{x}&=\frac{m\omega_{\text{eff}}^{x}}{2}\cot{{{\omega_{\text{eff}}^{x}}T}}[x_a^2+x_b^2\cos^2{\omega_{L}T}+y_b^2\sin^2{\omega_{L}T}-x_b y_b\sin{2\omega_{L}T}]\\
&-\frac{m\omega_{\text{eff}}^{x}}{\sin{{{\omega_{\text{eff}}^{x}}T}}}[x_a y_b\sin{\omega_{L}T}-x_a x_b\cos{\omega_{L}T}]\\
\end{split}
\end{equation}
\begin{equation}
\begin{split}
S_{\text{c}}^{y}&=\frac{m\omega_{\text{eff}}^{y}}{2}\cot{{{\omega_{\text{eff}}^{y}}T}}[y_a^2+x_b^2\sin^2{\omega_{L}T}+y_b^2\cos^2{\omega_{L}T}+x_b y_b\sin{2\omega_{L}T}]\\
&-\frac{m\omega_{\text{eff}}^{y}}{\sin{{{\omega_{\text{eff}}^{y}}T}}}[y_a y_b\cos{\omega_{L}T}+y_a x_b\sin{\omega_{L}T}]
\end{split}
\end{equation}
\end{subequations}
The result Eq.~(\ref{prop_anisotrop_ho}) has been achieved in $\text{Ref.}^{\cite{cervero:17}}$ by using the Method of the Stationary Phase Approximation, \ie \emph{Van Vleck - Pauli - Morette Determinant}. (See \eg Ref.\cite{blau:2014}) The author of $\text{Ref.}^{\cite{cervero:17}}$ obtains a fluctuation term analogous to Eq.~(\ref{prop_anisotrop_ho}). A quick numerical comparison shows that both expressions are identical. However, the identity is not manifestly evident. The result given in Eq.~(\ref{prop_anisotrop_ho}) is evidently shorter.
\subsubsection{Isotropic case $\omega_x=\omega_y$:}
Once again, a short calculation leads to the following rather expanded expression for the \emph{classical} action:
\begin{equation}
\begin{split}
S_{\text{c}}&=\frac{m\omega_{\text{eff}}}{2\sin{\omega_{\text{eff}}T}}\left[A_{1}\cos{\omega_{\text{eff}}T}-A_{2}\cos{\omega_{\text{L}}T}-A_{3}\sin{\omega_{\text{L}}T}\right]+A_{4}
\end{split}
\end{equation}
where
\begin{equation}
\begin{split}
A_1&=x_a^2+x_b^2+y_a^2+y_b^2+\frac{2q^2}{m^2\omega^4}\bold{E}^2-\frac{2q}{m\omega^2}[(x_a+x_b)E_x+(y_a+y_b)E_y]\\
A_2&=\frac{2q}{m\omega^2}[(x_a+x_b)E_x+(y_a+y_b)E_y]-\frac{2q^2}{m^2\omega^4}\bold{E}^2-2x_ax_b-2y_ay_b\\
A_3&=2x_ay_b-2x_b y_a-\frac{2q}{m\omega^2}[(y_b-y_a)E_x-(x_b-x_a)E_y]\\
A_4&=\frac{q^2 T}{2m\omega^2}\bold{E}^2
+\frac{q\omega_\text{L}E_x}{\omega^2}\left(y_b-y_a\right)-\frac{q\omega_\text{L}E_y}{\omega^2}\left(x_b-x_a\right)
\end{split}
\end{equation}
The formula for the \emph{fluctuation} integral is given by Eq.~(\ref{fluctuation_ho_b})
\section{Energy spectrum}
\label{spectrum}
In order to gain some confidence in the results achieved in the previous section, we calculate the energy spectrum of the charged \emph{isotropic} HO in crossed static electric and magnetic fields. The spectrum can be calculated directly from the propagator.
The corresponding \emph{anisotropic} case could be calculated in principle along the same track, albeit quite laboriously. 
By using the abbreviations $\bold{a}=(x,y)$, $\bold{E}=(E_x,E_y)$ and $\omega_{\text{\text{eff}}}=\sqrt{\omega^2+\omega_{L}^2}$ one has:
\begin{equation}
\label{mehler}
\begin{split}
&\text{Tr}\left(\exp\left[-\frac{i}{\hbar}H T\right]\right)=\int \limits_{\mathbb{R}^2}dx dy \, K(\bold{a},T;\bold{a},0)
\end{split}
\end{equation}
where Tr denotes the trace of the operator $\exp{\left[-\frac{i}{\hbar}HT\right]}$. The self-propagating Kernel $K$ is given by:
\begin{equation}
K(\bold{a},T;\bold{a},0)=\exp{\left[\frac{im\omega_{\text{eff}}}{\hbar}\frac{\cos{\omega_{\text{eff}}T}-\cos{\omega_{\text{L}}T}}{\sin{\omega_{\text{eff}}T}}D+\frac{iq^2T}{2m\hbar\omega^2}\bold{E}^2\right]}
\end{equation}
with 
\begin{equation}
D=\bold{a}^2-\frac{2q}{m\omega^2 }\bold{a}\cdot\bold{E}
\end{equation}
The Fresnel Integral \eqref{mehler} can easily be calculated. After a few calculational steps (including a quadratic completion) one obtains:
\begin{equation}
\begin{split}
&=\frac{1}{2}\, \frac{1}{{\cos{\omega_{\text{\text{eff}}}T}-\cos{\omega_{L}T}}}\exp{\left[\frac{iq^2T}{2m\hbar\omega^2}\bold{E}^2\right]}\\
&=\frac{1}{2i\sin(\frac{\omega_{\text{\text{eff}}}+\omega_{L}}{2}T)}\cdot\frac{1}{2i\sin(\frac{\omega_{\text{\text{eff}}}-\omega_{L}}{2}T)}\exp\left[\frac{iq^2T}{2m\hbar \omega^2}\bold{E}^2\right]\\
&=\exp\left[\frac{iq^2T}{2m\hbar \omega^2}\bold{E}^2\right]\sum_{n,m=0}^{\infty}\exp{\left[-i(\frac{\omega_{\text{\text{eff}}}+\omega_{L}}{2})T(n+\frac{1}{2})\right]}\exp{\left[-i(\frac{\omega_{\text{\text{eff}}}-\omega_{L}}{2})T(m+\frac{1}{2})\right]}\\
\end{split}
\end{equation}
By comparing the latter expression with the analogous one of a two-dimensional anisotropic Harmonic Oscillator one may read off the spectrum directly:
\begin{equation}
\begin{split}
E(n,m)&=\frac{q^2}{2m\omega^2}\bold{E}^2+\hbar(\sqrt{\omega^2+\omega_{L}^2}+\omega_{L})(n+\frac{1}{2}) \\
&+\hbar(\sqrt{\omega^2+\omega_{L}^2}-\omega_{L})(m+\frac{1}{2})
\end{split}
\end{equation}
In the limit $\omega\rightarrow0$ (and $\bold{E}=0$) one gets the expected \emph{Landau} levels\cite{landau:30,landau:1958} $E(n)=\hbar \omega_{c}(n+\frac{1}{2})$
where $\omega_{c}=\frac{qB}{m}$ denotes the cyclotron frequency.
\appendix
\section{Failure of the real Fourier sine expansion}
\label{failure_real_fourier}
An evaluation of path integrals by a Fourier sine series of the form Eq.~(\ref{real_expansion}) is a frequently used approach. (See \eg Ref.\cite{feynman:2010}) However, this method fails (as shown below) when magnetic fields are present.
This fact stands in contrast to an evaluation by the complex Fourier series Eq.~(\ref{fourier_exp}) where one can calculate the corresponding path integral correctly provided that one $\zeta$-regularizes the integral appropriately. 
\\
In terms of cartesian "coordinates" one has:
\begin{equation}
\begin{split}
K(\bold{b},T;\bold{a},0)
&=\int\mathcal{D}x \mathcal{D}y\,\exp{\left[\frac{im}{2\hbar}\int\limits_{0}^T \left(\dot{x}^2 + \dot{y}^2+\omega_c x\dot{y}-\omega_c y\dot{x}\right) dt\right]}
\end{split}
\end{equation}
By splitting:
\begin{equation}
\begin{split}
x(t)&=x_{\text{c}}+\delta x(t)\\
y(t)&=y_{\text{c}}+\delta y(t)
\end{split}
\end{equation}
one obtains:
\begin{equation}
\begin{split}
K(\bold{b},T;\bold{a},0)
&=\exp{\left[\frac{i}{\hbar}S_{\text{c}}\right]}\int\mathcal{D}[\delta x] \mathcal{D}[\delta y]\,\exp\left[{\frac{im}{2\hbar}\int\limits_{0}^T dt\left(\dot{\delta x}^2 + \dot{\delta y}^2+\omega_c\delta x\dot{\delta y}-\omega_c\delta y\dot{\delta x}\right)}\right]
\end{split}
\end{equation}
Expansions of quantum fluctuations $\delta x, \delta y$ in terms of eigenmodes  $\left(\phi_{n}\right)$ given in the introduction (See Eq.~(\ref{real_expansion}):
\begin{equation}
\delta x(t)=\sum\limits_{n=1}^{+\infty}a_{n} \phi_{n} \quad \delta y(t)=\sum\limits_{m=1}^{+\infty}b_{m} \phi_{m}
\end{equation}
are leading to:
\begin{equation}
\int\limits_{0}^T \delta x\dot{\delta y} dt = T \int\limits_{0}^T \langle \delta x|\dot{\delta y}\rangle dt =\sum\limits_{n, m=1}^{+\infty}a_{n} b_{m}\langle\phi_{n}|\dot{\phi}_{m}\rangle
\end{equation}
A short calculation shows:
\begin{equation}
\langle\phi_{n}|\dot{\phi}_{m}\rangle= 0  \quad \forall n, m \in \mathbb{N}
\end{equation}
thus:
\begin{equation}
\langle \delta x|\dot{\delta y}\rangle=\langle \delta y|\dot{\delta x}\rangle=0
\end{equation}
Therefore, one concludes that the propagator is identical to a free particle's one. This obviously erroneous result shows the failure of the applied method. 
\section{Remarks about convergence}
\label{remark_convergence}
The inner products ${\langle \dot{\delta \gamma}|\dot{\delta \gamma}\rangle}$ and ${\langle{\delta \gamma}|{\delta \gamma}\rangle}$ arising in the path integral \eqref{general_propagator} are undefinied as they stand. In the following calculation it is shown how this conceptional shortcoming can be overcome by introducing an exponential into the the Fourier series Eq.~(\ref{fourier_exp}).\\
By defining:
\begin{equation}
\label{eta_epsilon}
\delta \gamma_{\epsilon}(t)=\sum_{n=-\infty}^{+\infty}c_{n}\exp\left[-\epsilon n^2\right]\exp\left[i\frac{2\pi}{T}n\cdot t\right] ,\quad c_n
\in \mathbb{C}
\end{equation}
One can interpret $c_{n}\exp\left[-\epsilon n^2\right]\equiv \tilde{c}_{n}$ as the Fourier coefficients of the function $\delta \gamma_{\epsilon}(t)$. 
For a given set $(c_{n})_{n\in \mathbb{N}}\in l^{2}(\mathbb{C})$ (\ie space of square-summable complex sequences) the function $\delta \gamma_{\epsilon}(t)$ is smooth since its Fourier coefficients are exponentially decaying: $\tilde{c}_{n}\sim \exp\left[-\epsilon n^2\right] $. By using definition Eq.~(\ref{eta_epsilon}), the inner product
${\langle \dot{\delta \gamma_{\epsilon}}|\dot{\delta \gamma_{\epsilon}}\rangle}$  is now well-defined. In the following simple analysis, this method will be applied to the path integral of an harmonic oscillator.
The sligthly modified definition for the \emph{fluctuation} integral now reads:
\begin{equation}
\label{fluctuation_epsilon}
\begin{split}
F(T)&=\lim_{\epsilon\to 0}
\int\limits_{\delta \gamma_{\epsilon}(0)=0}^{\delta \gamma_{\epsilon}(T)=0}\mathcal{D}[\delta \gamma_{\epsilon}(t)]
\exp\left[\frac{imT}{2\hbar}\langle \dot{\delta \gamma_{\epsilon}}|\dot{\delta \gamma_{\epsilon}}
\rangle-\frac{im\omega^2 T}{2\hbar}\langle\delta \gamma_{\epsilon}|\delta \gamma_{\epsilon}\rangle\right]
\end{split}
\end{equation}
By exactly the same procedure as applied in Eq.~(\ref{fluctuation_1d}) one obtains the \emph{fluctuation} integral in one dimension:
\begin{equation}
F_{1\text{d}}(T)=\lim_{\epsilon\to 0}\int\prod_{n=-\infty}^{+\infty}dx_{n}
\exp{\left[i \sum_{n=-\infty}^{+\infty}|x_{n}|^2 \left(\alpha n^2-\beta\right)\exp[-2\epsilon n^2]\right]}
\delta\left(\sum\limits_{n=-\infty}^{+\infty}x_n \exp[-\epsilon n^2]\right)
\end{equation}
After a few simple calculational steps one gets:
\begin{equation}
\begin{split}
F_{1\text{d}}(T)&=\frac{1}{2\pi}\lim_{\epsilon\to 0}\int dk \prod_{n=-\infty}^{+\infty} \exp\left[\frac{-ik^2}{4(\alpha n^2-\beta)}\right]\sqrt{\frac{i\pi \exp[2\epsilon n^2]}{\alpha n^2-\beta}}\\
&=\frac{1}{2\pi}\lim_{\epsilon\to 0}\int dk \prod_{n=-\infty}^{+\infty} \exp\left[\frac{-ik^2}{4(\alpha n^2-\beta)}\right]\sqrt{\frac{i\pi}{\alpha n^2-\beta}}
\prod_{n=1}^{+\infty} \exp[-2\epsilon n^2]
\end{split} 
\end{equation}
Since 
\begin{equation}
\prod_{n=1}^{+\infty} \exp\left[-2\epsilon n^2\right]=\exp[-2\epsilon\underbrace{\zeta(-2)}_{=0}]=1
\end{equation}
one gets exactly the same expression as in Eq.~(\ref{calculation}), even without performing the limit $\epsilon \rightarrow 0$. This means that (path-)integration over smooth functions defined by Eq.~(\ref{eta_epsilon}) gives the same result as an integration over space $L^2_{\mathbb{C}}[0,T]$, provided one $\zeta$-regulates the corresponding integrals.
\section{Heuristic derivation of some values of $\bold{\zeta(s)}$ and $\bold{\zeta^{\prime}(0)}$}
\label{values_zeta}
This section is entirely expository. The values of $\zeta(s)$ for some $s\in \mathbb{Z}$ and $\zeta^{\prime}(0)$ will be calculated in a refreshing but non-rigorous manner. A mathematically correct derivation would require Riemann's well-known functional equation for $\zeta(s)$. (See \eg Ref.\cite{titchmarsh:1951}) Nevertheless, it is interesting to look at these special values in the broader context of diverging sums which are interesting mathematical objects in its own right. (see \eg Ref.\cite{hardy:73}). 
\subsection{Values of $\bold{\zeta(s)}$, $s\bold{\in \mathbb{Z}}$}
In order to calculate the values $\zeta(s)$ for $s\in \mathbb{Z} $ one considers a class of alternating and divergent series formally given by:
\begin{equation}
\label{alter_sum}
\eta(-s):=\sum\limits_{n=1}^{+\infty}(-1)^{n+1} n^s ,\quad s\in \mathbb{Z}
\end{equation}
A few examples:
\begin{equation}
\begin{split}
s=1: \quad &1-2+3-4+5-6+\cdots  \\
s=2: \quad&1^2-2^2+3^2-4^2+5^2-6^2+\cdots \\
s=3: \quad&1^3-2^3+3^3-4^3+5^3-6^3+\cdots\\ 
\end{split}
\end{equation}
These series are divergent according to Cauchy's orthodoxy. However, by inserting an exponential into the series \eqref{alter_sum} they turn into convergent ones:
\begin{equation}
\label{sum_converge}
\eta_{\epsilon}(-s):=\sum\limits_{n=1}^{+\infty}(-1)^{n+1} n^s \mathrm{e}^{-\epsilon n}
\end{equation}
By taking the limit $\epsilon\to 0$ one attributes a well-defined value to the divergent series \eqref{alter_sum}:
\begin{equation}
\label{sum_converge}
\lim_{\epsilon\to 0}\eta_{\epsilon}(-s):=\lim_{\epsilon\to 0}\sum\limits_{n=1}^{+\infty}(-1)^{n+1} n^s \mathrm{e}^{-\epsilon n}
\end{equation}
The latter series can be evaluated explicitly (by using the formula for an infinite geometric series):
\begin{equation}
\label{formula_sum}
\eta(-s):=(-1)^{s} \lim_{\epsilon\to 0}\left( \frac{d^s}{d\epsilon^s}\frac{1}{1+\mathrm{e}^{\epsilon}}\right)
\end{equation}
The latter formula enables to calculate at least some of the series \eqref{alter_sum}:
\begin{equation}
\begin{split}
\eta(0)&=1-1+1-1+\cdots=\frac{1}{2}\\
\eta(1)&=1-2+3-4+\cdots= \frac{1}{4}\\
\end{split}
\end{equation}
For $s=-1$:
\begin{equation}
\begin{split}
&1-\frac{1}{2}+\frac{1}{3}-\frac{1}{4}+\cdots=(-1)^{-1}\lim_{\epsilon\to 0}\,\left(\frac{d^{-1}}{d\epsilon^{-1}}\frac{1}{1+\mathrm{e}^\epsilon}\right)\\
&=-\lim_{\epsilon\to 0}\int\frac{1}{1+e^\epsilon}d\epsilon=(-1)^{-1}\lim_{\epsilon\to 0}\left[-\log|1+e^{-\epsilon}|+C\right]
\end{split}
\end{equation}
The integration constant $C$ must vanish since $ \lim_{\epsilon\to \infty}\eta_{\epsilon}(1)=0$.
One has thus the expected value:
\begin{equation}
\eta(1)=1-\frac{1}{2}+\frac{1}{3}-\frac{1}{4}+\cdots=\log 2
\end{equation}
One can easily derive an analogous formula for the corresponding non-alternating series which is nothing but an expression for the Riemann-$\zeta$-function for (at least) $s\in \mathbb{Z}$:
\begin{equation}
\zeta(-s)=\sum\limits_{n=1}^{+\infty}n^s=\frac{(-1)^{s}}{1-2^{1+s}} \lim_{\epsilon\to 0} \frac{d^s}{d\epsilon^s}\left(\frac{1}{1+\mathrm{e}^{\epsilon}}\right) ,\quad s\in \mathbb{Z}
\end{equation}
Once again a few examples:
\begin{equation}
\begin{split}
\zeta(0)&=1+1+1+1+\cdots = -\frac{1}{2}\\
\zeta(-1)&=1+2+3+4+\cdots = -\frac{1}{12}\\
\zeta(-2)&=1+4+9+16+\cdots=0
\end{split}
\end{equation}
\subsection{Derivation of $\bold{{\zeta^{\prime}(0)}}$}
In order to calculate ${\zeta^{\prime}(0)}$ one considers the following (divergent) series:
\begin{equation}
\label{alternating_sum_logarithm}
\sum\limits_{n=1}^{+\infty}(-1)^{n+1}\log n
\end{equation}
The latter series can be calculated formally (heuristically) by using Wallis' product\cite{wallis:1656} for $\pi$. One starts with the identity: 
\begin{equation}
\label{wallis}
\frac{\pi}{2}=\frac{2}{1}\cdot\frac{2}{3}\cdot\frac{4}{3}\cdot\frac{4}{5}\cdot\frac{6}{5}\cdot\frac{6}{7}\cdots
\end{equation}
By handwaving one gets:
\begin{equation}
\label{wallis_modified}
\begin{split}
\Longrightarrow \sqrt{\frac{\pi}{2}}&=\frac{2\cdot4\cdot6\cdot8\cdots}{1\cdot3\cdot5\cdot7\cdots}\\
\Longrightarrow \sqrt{\frac{2}{\pi}}&=\frac{1}{2}\cdot\frac{3}{4}\cdot\frac{5}{6}\cdot\frac{7}{8}\cdots
\end{split}
\end{equation}
On the other hand one has:\\
\begin{equation}
\label{wallis_2}
\begin{split}
\sum\limits_{n=1}^{+\infty}(-1)^{n+1}\log n&=\log1-\log2+\log3-\log4+\cdots\\
&=\log\frac{1}{2}+\log\frac{3}{4}+\log\frac{5}{6}+\cdots\\
&=\log(\frac{1}{2}\cdot\frac{3}{4}\cdot\frac{5}{6}\cdot\frac{7}{8}\cdots)
\end{split}
\end{equation}
Substitution of Eq.~(\ref{wallis_modified}) into Eq.~(\ref{wallis_2}) yields:\\
\begin{equation}
\label{sum_log_alt}
\sum\limits_{n=1}^{+\infty}(-1)^{n+1}\log n=\log\sqrt{\frac{2}{\pi}}=\frac{1}{2}\log\frac{2}{\pi}
\end{equation}
The non-alternating series  $X:=\sum\limits_{n=1}^{+\infty}\log n$ can also be evaluated in a formal manner. By adding another (divergent) series to \eqref{alternating_sum_logarithm} one gets:
\begin{equation}
\label{add_sum}
\begin{split}
&\log1-\log2+\log3-\log4+\log5-\log6\cdots\\
&+2\log2    +2\log4   +2\log6 \cdots=X
\end{split}
\end{equation}
Considering that\\
\begin{equation}
\begin{split}
&2\log2    +2\log4   +2\log6+ \cdots\\
&=2(\log1+\log2+\log2+\log2+\log2+\log3+\cdots)\\
&=2\log2\underbrace{\left(\sum\limits_{n=1}^{+\infty}1\right)}_{=-\frac{1}{2}}+2(\log1+\log2+\log3+\log4+\log5+\log6\cdots)\\
&=-\log2+2X
\end{split}
\end{equation}
Rewriting Eq.~(\ref{add_sum}) in the form:
\begin{equation}
\begin{split}
X=2X-\log2+\frac{1}{2}\log\frac{2}{\pi}
\end{split}
\end{equation}
Solving for $X$:
\begin{equation}
X=\sum\limits_{n=1}^{+\infty}\log n=\frac{1}{2}\log2\pi
\end{equation}
Thus:
\begin{equation}
\zeta^{\prime}(0)=-\sum\limits_{n=1}^{+\infty}\log n=-\frac{1}{2}\log2\pi
\end{equation}

\begin{acknowledgments}
The author would like to thank J\"urgen Kies for interesting discussions, which inspired this paper.
\end{acknowledgments} 
\appendix*

\bibliography{paper_path_integral_11_08_19}

\begin{thebibliography}{10}

\bibitem{ahlfors:78}
L.~V. Ahlfors.
\newblock {\em Complex analysis, An introduction to the theory of analytic
  functions of one complex variable, third ed.}
\newblock McGraw-Hill Book Co., New York, 1978, 1978.

\bibitem{blau:2014}
M.~Blau.
\newblock {\em http://www.blau.itp.unibe.ch/lecturesPI.pdf)}.
\newblock 2014.

\bibitem{cameron:60}
R.H. Cameron.
\newblock A family of integrals serving to connect the wiener and feynman
  integrals.
\newblock {\em Journal of Math. and Phys}, 39:126--140, 1960.

\bibitem{cervero:17}
J.M. Cervero.
\newblock Exact propagator of a two dimensional anisotropic harmonic oscillator
  in the presence of a magnetic field.
\newblock {\em Journal of Modern Physics}, 8:500--510, 2017.

\bibitem{courant:43}
R.~Courant and D.~Hilbert.
\newblock {\em Methoden der mathematischen Physik, 2nd ed.}, volume~1.
\newblock Interscience Publishers, Inc. New York, 1943.

\bibitem{dirac:1930}
P.~A.~M. Dirac.
\newblock {\em Principles of Quantum Mechanics}.
\newblock Oxford University Press, 1930.

\bibitem{feynman:48}
R.~P. Feynman.
\newblock Space-time approach to nonrelativistic quantum mechanics.
\newblock {\em Rev.\ Mod.\ Phys.}, 20:367--387, 1948.

\bibitem{feynman:2010}
R.~P. Feynman and A.~R. Hibbs.
\newblock {\em Quantum Mechanics and Path Integrals}.
\newblock Dover Publications, New York, 2010.

\bibitem{gelfand:60}
I.M. Gelfand and A.M. Yaglom.
\newblock Integration in functions spaces and its applications in quantum
  physics.
\newblock {\em J. Math. Phys. 1}, 1:48--69, 1960.

\bibitem{gradshteyn:2014}
I.S. Gradshteyn and I.M. Ryzhik.
\newblock {\em Table of Integrals, Series and Products, eighth ed.}
\newblock Academic Press, 2014.

\bibitem{hardy:73}
G.H. Hardy.
\newblock {\em Divergent Series}.
\newblock Oxford at the Clarendon Press, 1973.

\bibitem{hurwitz:1902}
A.~Hurwitz.
\newblock quelques applications géometrique des séries de fourier.
\newblock {\em Annales de l'Ecole Normale}, 19:357--408, 1902.

\bibitem{kac:51}
M.~Kac.
\newblock On some connection between probability theory and differential and
  integral equations.
\newblock {\em Proc, 2nd Berkeley Sympos. Math. Stat. and Prob.}, pages
  189--215, 1951.

\bibitem{kleinert:2004}
H.~Kleinert.
\newblock {\em Path Integrals in Quantum Mechanics, Statistics, Polymer
  Physics, and Financial Markets}.
\newblock World Scientific, Singapore, 2004.

\bibitem{kline:1972}
M.~Kline.
\newblock {\em Mathematical Thought from Ancient to Modern Times}, volume~1.
\newblock Oxford University Press, 1972.

\bibitem{landau:30}
L.~D. Landau.
\newblock L. landau, z. physik (1930) 64: 629.
\newblock {\em Z. Physik}, 64:629, 1930.

\bibitem{landau:1958}
L.~D. Landau and E.~M. Lifshitz.
\newblock {\em Quantum Mechanics, Non-Relativistic Theory}, volume~3.
\newblock Pergamon Press, Oxford, 1958.

\bibitem{leibniz:1684}
G.W. Leibniz.
\newblock {\em Nova Methodus pro Maximis et Minimis, Acta Eruditorum}.
\newblock 1684.

\bibitem{mazzucchi:09}
S.~Mazzucchi.
\newblock {\em Mathematical Feynman Path Integrals And Their Applications}.
\newblock World Scientific, 2009.

\bibitem{montroll:52}
E.W. Montroll.
\newblock Markov chains, wiener integrals and quantum theory.
\newblock {\em Comm. Pure Appl. Math.}, 5:415, 1952.

\bibitem{schulman:1981}
L.~S. Schulman.
\newblock {\em Techniques and Applications of Path Integration}.
\newblock John Wiley and Sons Inc., New York, 1981.

\bibitem{titchmarsh:1951}
E.~C. Titchmarsh.
\newblock {\em The theory of the Riemann zeta-function}.
\newblock Oxford University Press, 1951.

\bibitem{wallis:1656}
J.~Wallis.
\newblock {\em Arithmetica Infinitorum, Opera I, p. 468}.
\newblock 1655.

\bibitem{werner:2011}
D.~Werner.
\newblock {\em Funktionalanalysis}.
\newblock Springer, Berlin, 2011.

\end{thebibliography}
\bibliographystyle{plain}
\end{document}